\documentclass[12pt,twoside,letterpaper,twocolumn]{article}
\usepackage[margin=20mm]{geometry}
\usepackage{amssymb,amsmath}
\usepackage{url}
\usepackage{graphicx}
\usepackage{appendix}

\newif\ifnature
\naturefalse

\newif\ifappendixtitle
\appendixtitlefalse

\begin{document}
\author{Robert M Brady\\
				\small University of Cambridge Computer Laboratory\\
        \small JJ Thomson Avenue, Cambridge CB3 0FD, United Kingdom\\
        \small \texttt{robert.brady@cl.cam.ac.uk}}

\ifappendixtitle
	\title{Supplementary Information\\The irrotational motion of a compressible inviscid fluid}
\else		
	\title{The irrotational motion of a compressible inviscid fluid}
\fi

\date{\today}

\twocolumn[
  \begin{@twocolumnfalse}
    \maketitle
    \begin{abstract}

		\ifnature
			The irrotational motion of a compressible inviscid fluid is studied in the field of analogue gravity, where its metric is compared to that in general relativity, a fluid analogue of an evaporating black hole has been realized experimentally, and there are symmetries related to the standard model. Here we show the analogy also extends quantitatively to electromagnetic and quantum mechanical phenomena. We discuss a candidate model to account for the number and precision of these analogies.
		\else
			The irrotational motion of a compressible inviscid fluid is studied in the field of analogue gravity, where its metric is compared to that in general relativity~\cite{barcelo2011analogue}, a fluid analogue of an evaporating black hole~\cite{unruh1981experimental} has been realized experimentally~\cite{lahav2010realization}, and there are symmetries related to the standard model~\cite{Volovik-2003}. Here we show the analogy also extends quantitatively to electromagnetic and quantum mechanical phenomena. We discuss a candidate model to account for the number and precision of these analogies.
		\fi
  	\end{abstract}
  \end{@twocolumnfalse}
]

%                              ----------------- Introduction -----------------

\section{Introduction}

We will show that Euler's equation for a compressible fluid has irrotational solutions which superficially resemble smoke rings. Their equations of motion through the fluid are completely classical, and the resulting trajectories obey statistical equations which are identical to those of the Copenhagen interpretation of quantum mechanics. A similar mechanism accounts for the emergence of quantum phenomena from classical motion in experiments where a droplet is made to bounce on a liquid surface, exhibiting quantised energy levels, single-slit diffraction, double-slit diffraction and tunnelling~\cite{couder2005dynamical,couder2006single,protiere2006particle,eddi2009unpredictable,fort2010path}. Some of the ring-like solutions are chiral. Opposite chiralities attract and like chiralities repel with an inverse square force which has the same fluid dynamic origin as the effect used for degassing oil by subjection to ultrasonic vibration~\cite[\S 4.4]{Faber}. The interaction obeys Maxwell's equations and its strength is characterized by a fine structure constant, $\alpha \lesssim 1/49$.

Consider a compressible inviscid fluid such as the air in the idealization it is continuous and has no viscosity or thermal conductivity. Such a fluid obeys Euler's equation~\cite{Faber}
\begin{equation}
	\frac{\partial {\bf u}}{\partial t} + (\textbf{u}. \nabla)\textbf{u} ~ = ~ - \frac{1}{\rho}\nabla P 
	\label{eq:euler}
\end{equation}
where the pressure $P(\textbf{x}, t)$ is a function of the density $\rho(\textbf{x}, t)$ and ${\partial \rho}/{\partial t} = -\nabla (\rho  \textbf{u})$. At low amplitude this reduces to the wave equation
\begin{equation}
	\frac{\partial^2 \rho}{\partial t^2} ~ - ~ c^2\nabla^2 \rho ~~ = ~~ 0
	\label{eq:wave}
\end{equation}
where $c^2 = \frac{\partial P}{\partial \rho}$. In addition to describing sound waves, this equation has cylindrical solutions, the simplest of which is present in the surface waves just after a raindrop has hit a puddle. In cylindrical coordinates $(r, \theta, s)$
\begin{equation}
	\xi ~ = A ~ e^{-i(\omega_o t + m \theta - k_s s)} ~J_m (k_r r)
	\label{eq:eddy}
\end{equation}
where $\frac{\rho}{\rho_o} = 1 + Re(\xi)$ and $\rho_o$ is the mean density, $J_m$ is a cylindrical Bessel function of the first kind, $m$ an integer and $\omega_o^{2} = c^2 (k_s^2 + k_r^2)$.  It may be verified by substitution that this obeys the wave equation. The flow pattern is irrotational, that is, $\oint {\bf u}.{\bf dl} = 0$ for any closed path, as shown in the appendix.

\begin{figure}[htb]
	\centering
	\includegraphics[width=0.3\textwidth]{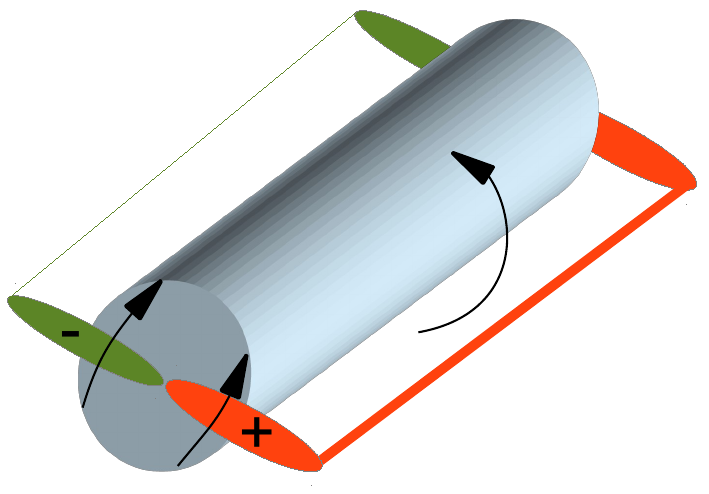}
	\caption{Sketch of an eddy, described by \eqref{eq:eddy} with $m = 1$ and $k_s = 0$. The compressions (red, schematic and not to scale) and rarefactions (green) rotate as shown. Note the contrary direction of fluid flow near the rarefaction.}
	\label{fig:eddy}
\end{figure}
\begin{figure}[htbp]
  \centering
	\includegraphics[width=\columnwidth]{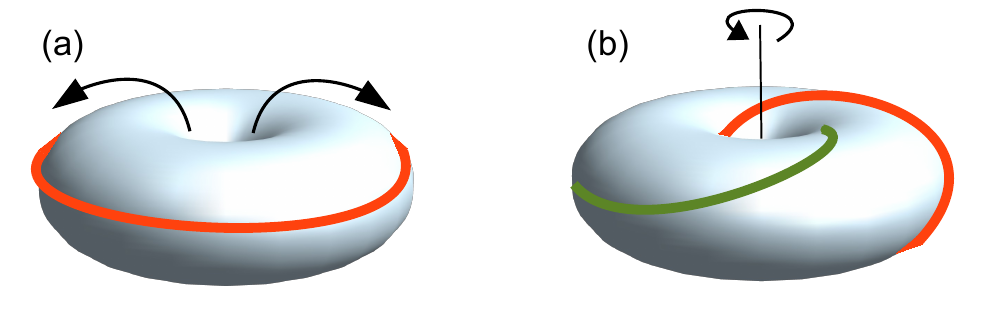}
	\caption{Two sonons obtained by curving the eddy of Figure \ref{fig:eddy} into a torus. The compressions (red) and rarefactions (green) rotate as shown. (a) $R_{10}$ (see equation \ref{eq:rmn}) (b) $R_{11}$, obtained by adding a twist. This solution is chiral and the motion resolves into a spin about the axis.}
	\label{fig:two-sonons}
\end{figure}

Figure \ref{fig:eddy} sketches the $m=1$ solution. There are families of solutions, superficially resembling smoke rings, which can be obtained by curving it into a torus. Two such solutions, which we call `sonons', are sketched in Figure \ref{fig:two-sonons}.

In the case $m = 0$ or $1$, the Appendix shows that the associated density pattern can be written to a good approximation as a sum of spherical harmonic solutions to the wave equation in the coordinates of figure \ref{fig:sonon-integration-path}:-
\begin{equation}
	\xi =\psi_o(t) ~ R_{mn}({\bf x})
\label{eq:factorise-xi}
\end{equation}
where
\begin{equation}
	\psi_o ~~=~~ A ~e^{-i \omega_o t}
	\label{eq:psi-o}
\end{equation}
\begin{equation}
	R_{mn} = \int_0^{2 \pi} e^{-i (m \theta' - n \phi)} j_m (k_r \sigma) k_r R_o d\phi
	\label{eq:rmn}
\end{equation}
where $n$ is an integer and $j_m$ is a spherical Bessel function of the first kind.

\begin{figure}[htbp]
  \centering
	\includegraphics[width=0.33\columnwidth]{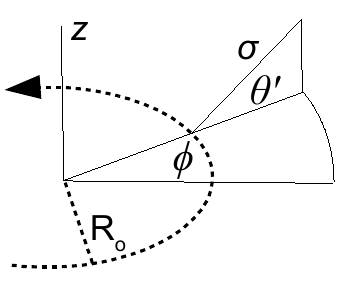}
	\caption{The coordinates used in (\ref{eq:rmn})}
	\label{fig:sonon-integration-path}
\end{figure}

\subsection{Dimensions of a sonon}
\label{ss:sonon-radius}

If the eddy in Figure \ref{fig:eddy} is terminated on two planar surfaces, they will attract one another due to the reduced Bernoulli pressure. This must be due to an attraction along the length of the eddy, similar to the attraction in a vortex.

When an ordinary vortex is curved into a smoke ring, this force is balanced by Magnus forces (like the lift of an aircraft wing) as the structure moves forward through the fluid~\cite{Faber}. However a sonon cannot experience Magnus forces because it is irrotational, and consequently its radius will shrink, causing the amplitude $A$ in \eqref{eq:psi-o} to grow due to the conservation of fluid energy. Nonlinear effects will halt the shrinking before $A$ reaches about 1 since the density cannot become negative. 
\begin{figure}[htb]
	\centering
		\includegraphics[width=\columnwidth]{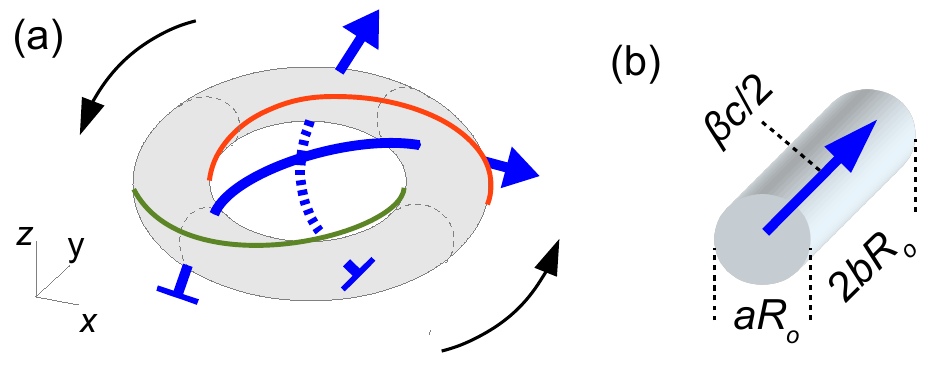}
	\caption{(a) The fluid flows (blue) near a $R_{11}$ sonon reinforce in the $y$ direction at the instant shown. (b) The nonlinear region of reinforcement approximates to a cylinder. If it has diameter is $a R_o$, length $2 b R_o$ and mean flow speed $\frac{1}{2}\beta c$, a reasonable estimate is $a \approx b \approx \beta \approx 1$}
	\label{fig:r11fluidflow-with-cylinder}
\end{figure}

Very near a $R_{11}$ sonon, shown in Figure \ref{fig:r11fluidflow-with-cylinder}, the main component of the flow derives from the radial flow speed due to an eddy with $m = 1$, which is $u_r \propto \frac{\partial}{\partial r} J_1(k_r r)$ (see the appendix). These flows reinforce provided the diameter of the sonon is less than the first zero of $u_r$, which is $2 \times 0.92$. Neglecting curvature and nonlinearities, the lowest energy (unexcited) state has, as an upper estimate,
\begin{equation}
	R_o ~~\lesssim~~ \frac{0.92}{k_r}
	\label{eq:R11-radius}
\end{equation}

\subsection{Spin symmetry}

The integral in \eqref{eq:rmn} can be factorised into 
$R_{mn}=[B_r.\Phi_r(r)] [B_\theta.\Phi_\theta(\theta)] [B_\phi.\Phi_\phi(\phi)] $,
where $B_i$ and $\Phi_i$ are vectors. 

At large distance, the components of $\Phi_r$ can be obtained by expanding the Bessel function at large $r$ and neglecting small terms
\begin{equation}
		\Phi_r ~~=~~ \frac{1}{r} ~ [~\sin(k_r r), ~\cos(k_r r)~]
\label{eq:phi-r}
\end{equation}
Notice that
\[
	i 
	\left(
		\begin{array}{cc}
			0&-i \\
			i&0
		\end{array}
	\right)
	\frac{\partial \Phi}{\partial r}=k_r \Phi
\]
where we have defined $\Phi = \Phi_r \Phi_\theta \Phi_\phi$. The other eigenvector, $[\sin(k_r r), -\cos(k_r r)]$, has an eigenvalue of the opposite sign.

By inspection of \eqref{eq:rmn} there is an eigenvector $\Phi_\phi = [e^{-i n \phi}, e^{i n \phi}]$, which obeys
\[
	i
	\left(
		\begin{array}{cc}
			1&0 \\
			0&-1
		\end{array}
	\right)
	\frac{\partial \Phi}{\partial \phi} =n \Phi
\]
Reversing the sign of $n$ corresponds to reversing the spin direction, and it produces the other eigenvector with the opposite sign of eigenvalue. 

In the $\theta$ direction, the contributions from the near and far sides of the sonon produce wave patterns which propagate in opposite directions at the calculation position in Figure \ref{fig:sonon-integration-path}. The two orthogonal components are $\Phi_\phi = [\Phi^+, \Phi^-]$ where $\Phi^\pm = e^{-i m \theta} \pm e^{i m \theta}$. Thus
\[
	i
	\left(
		\begin{array}{cc}
			0&1 \\
			1&0
		\end{array}
	\right)
	\frac{\partial \Phi}{\partial \theta} = m \Phi
\]
Notice that these matrices are the same as the Pauli spin matrices. The appendix discusses a possible physical interpretation of the associated spin-half symmetry, and also the analogy with the Dirac equation. 

\subsection{Boundary condition}
\label{ss:standing-waves}

Taking one of the components of $\Phi_r$ in \eqref{eq:phi-r}, the kinetic energy $\frac{1}{2} \int \rho u^2$ can be calculated in the usual way in fluid dynamics. The value must be doubled to account for the potential energy of motion. The total energy inside a sphere of radius R is (see the appendix)
\begin{equation}
	E_{fluid} ~~ \approx ~~ \pi R_o^2 ~ R~ \rho_o c^2
\label{eq:sonon-energy-r}
\end{equation}

This approximates to the energy in the fluid motion of sonon since $A$ and $B_{ri}$ are of order 1 and the others diminish rapidly with radius.

Even without container walls, there will still be a boundary condition due to other sonons, since the energy in (\ref{eq:sonon-energy-r}) will be reduced if they align themselves so they interfere destructively at large distance. The coupling mechanism which allows the lowest energy state to be reached was first reported in pendulum clocks by Huygens in 1665~\cite{bennett2002huygens} -- see the appendix.

Figure \ref{fig:spin-align} shows two $R_{11}$ sonons which are aligned so they interfere destructively at large distance. The spherical Bessel function $j_1(r)$ changes phase by $\pi$ compared to a freely propagating wave when going through the centre, and it follows that destructive interference at large distance means there must be constructive interference in the region between the sonons. There is consequently an antinode of density on the mirror line. Notice that the spin vectors in Figure \ref{fig:spin-align} are opposed. 

\begin{figure}[htb]
	\centering
		\includegraphics[width=.95\columnwidth]{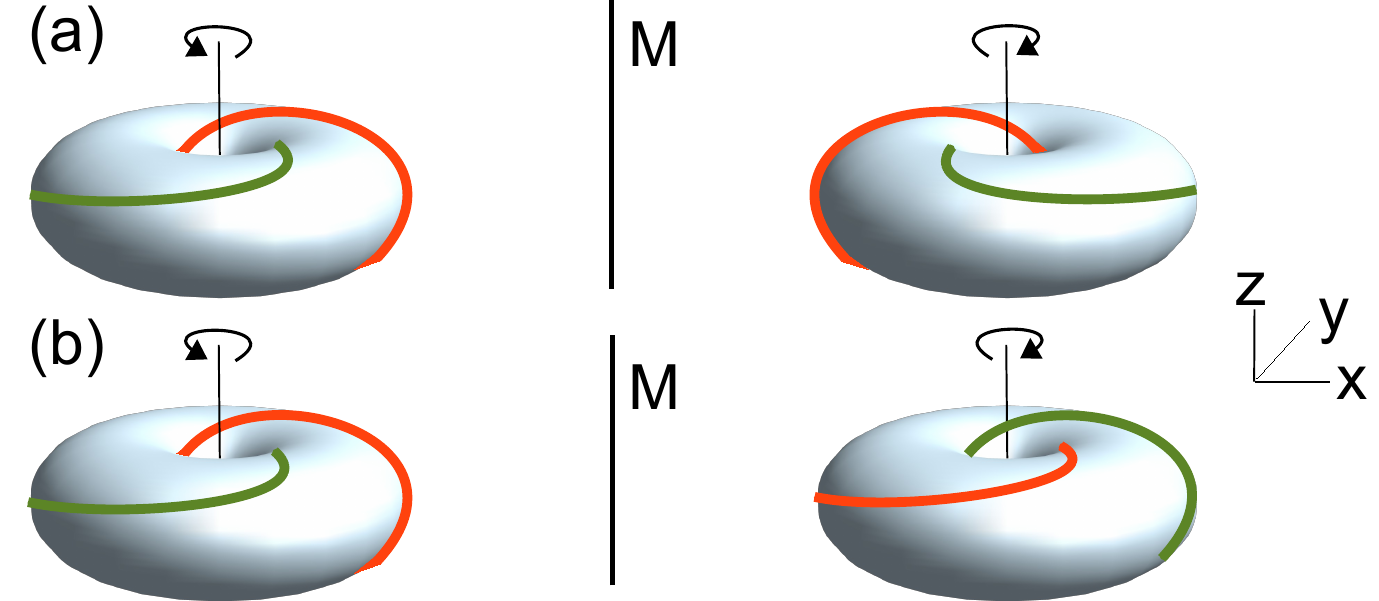}	\caption{Pairs of $R_{11}$ sonons of (a) opposite (b) the same chirality, aligned so their waves interfere destructively at large distance on an extended line joining them. There is an antinode of density on the mirror line $M$.}
	\label{fig:spin-align}
\end{figure}

Suppose the sonons in a region are brought closer together to form a denser body where the typical distance $R$ between them is reduced. If the process is adiabatic then, from \eqref{eq:sonon-energy-r}, $RR_o^2$ will remain constant, and consequently the sonons will become enlarged, with a greater value of $R_o$. These enlarged structures are valid solutions to Euler's equation provided their frequencies are also redshifted, since if $\xi({\bf x}, t)$ is a solution then it may be verified by substitution into \eqref{eq:euler} that $\xi(a {\bf x}, a t)$ is also a solution where $a$ is an arbitrary scale factor (noting that the velocity $u$ is unaffected by the scaling).  A similar redshift near a massive body is studied in analogue gravity, as discussed further in the appendix.

%                                ------------------ Special relativity --------------
\section{Lorentz covariance}
\label{sec:lorentz-covariance}

From \eqref{eq:sonon-energy-r}, most of the fluid energy of a sonon is at large distance, where the motion obeys the wave equation because it is low amplitude. Since the wave equation is Lorentz covariant, so is this motion. Thus, if $\xi({\bf x}, t)$ is a solution then $\xi({\bf x}', t')$ is also a solution where $x'=\gamma(x-vt)$, $y'=y$, $z'=z$, $t'=\gamma(t-v x/c^2)$ and $\frac{1}{\gamma} = \sqrt{1-v^2/c^2}$.

Very near the sonon, the amplitude of motion is not negligible and the Lorentz symmetry might, in principle, be perturbed by the nonlinear term $(\textbf{u}.\nabla) \textbf{u}$ in (\ref{eq:euler}). However, consider adding a constant velocity to all motion. Replacing $\textbf{u}$ by $\textbf{u} + \textbf{v}$ where $\bf v$ is constant gives a perturbation $\epsilon = (\textbf{v}.\nabla) \textbf{u}$. This perturbation vanishes over a cycle because the fluid motion is oscillatory and returns to its starting-point, $\int \textbf{u}~dt = 0$ and so 
\[
	\int \epsilon ~dt ~~=~~ \int (\textbf{v}.\nabla) \textbf{u} ~dt ~~=~~ 0
\]
Expectation values are usually defined to converge on the long term measurement. For precision we define an `ordinary' expectation value where this measurement encompasses one or more complete cycles. It follows that no deviations from Lorentz covariance can be discovered by measuring any ordinary expectation value.

Another potential perturbation arises because the speed $c = \sqrt{\partial P/\partial \rho}$ varies over the cycle. However, in the absence of external influences, the speed of a propagating wave is unperturbed because its leading edge cannot travel faster than the low amplitude speed, as is known for sound waves which obey similar equations~\cite{Faber}.

Finally, a sonon might be perturbed by the mean variation in $c^2 = \frac{\partial P}{\partial \rho}$. However, the effect is third-order, as can be seen by expanding $P(\rho)$ in terms of $s = (\rho - \rho_o)/\rho_o$: the mean of $\frac{\partial P}{\partial \rho}$ is only perturbed by the odd-order terms, starting with $s^3$. Even if the effect were to rise above noise levels, we will see it would be manifested as an adjustment to the effective mass of the quasiparticle, which would be difficult to distinguish from other effects such as thermal energy.

It follows that all ordinary expectation values are Lorentz covariant to a very good approximation at all amplitudes.

\section{Equations of motion}

From \eqref{eq:psi-o}, $\nabla^2 \psi_o = 0$ and $\frac{\partial^2}{\partial t^2} \psi_o = -\omega_o^2 \psi_o$. We have seen the sonon is Lorentz covariant to a very good approximation, so this must be a special case of a Lorentz covariant relationship which is velocity-independent, namely
\begin{equation}
	\frac{\partial^2 \psi}{\partial t^2}   ~ - ~ c^2 \nabla^2 \psi ~~ = ~~ -\omega_o^2 \psi
\label{eq:klein-gordon}
\end{equation}
which is analogous to the Klein-Gordon equation for a relativistic particle. At low velocity, \eqref{eq:klein-gordon} can be approximated by defining $\psi = e^{-i \omega_o t} \psi'$ and neglecting $\frac{\partial^2}{\partial t^2} \psi'$, giving
\begin{equation}
	i \hbar \frac{\partial \psi'}{\partial t} + \frac{\hbar^2}{2 m} \nabla^2 \psi' ~~= ~~ V \psi'
\label{eq:schrodinger}
\end{equation}
where, in a change of units to match convention, we have defined $m = \frac{\hbar \omega}{c^2}$ and $V$ is a constant. This is analogous to the Schr\"odinger equation for motion in a constant potential. 

To complete the description, applying a Lorentz boost to the density pattern $R_{mn}$ in \eqref{eq:rmn} and neglecting its time dependence (which is second order in $\bf v$) gives the same equation with $r$ replaced by $r'$ where $r_x' = \gamma (r_x - v t)$, $r_y' = r_y$ and $r_z' = r_z$. We will refer to this as $\chi({\bf x})$. Its peaks and troughs also move through space at the velocity $\bf v$, and consequently the sonon will remain aligned with them if it is unperturbed. This coherence is likely to be reinforced by nonlinearities in the medium. We will first examine the coherent case, and then examine the effects of decoherence.

An expression for the velocity $\bf v$ in terms of $\psi$ follows by applying a Lorentz boost to \eqref{eq:psi-o}, giving $\psi = A e^{i S}$ where $S = {\bf k.x}-\omega t$ and, from the Lorentz transformation, $\omega = \gamma \omega_o$ and ${\bf k}=(\gamma \omega_o/c^2) {\bf v}$. These combine to $\hbar {\bf k} = m {\bf v}$ where we have defined $m = \frac{\hbar \omega}{c^2}$. This leads immediately to
\begin{equation}
	{\bf v} ~= ~\frac{\hbar}{m} ~Im \left( \frac{\nabla \psi}{\psi} \right)
\label{eq:bohm-velocity}
\end{equation}
The significance of this expression can be understood by following Bohm~\cite{bohm1952suggested}, who rearranged the Shr\"odinger equation into the form
\begin{equation}
	\frac{\partial |\psi|^2}{\partial t} ~~+~~ \nabla(|\psi|^2 {\bf v})~~=~~0
\label{eq:bohm-probability}
\end{equation}
where $\bf v$ is given by \eqref{eq:bohm-velocity}. Bohm noticed that \eqref{eq:bohm-probability} is the standard equation for a conserved quantity moving at velocity $\bf v$, and concluded that $|\psi|^2$ must be the probability (averaged over nearby trajectories) that a hypothetical particle moving at velocity $\bf v$ will pass through a given position at a given time. These particle trajectories are the same as those of sonons. However the resulting de Broglie-Bohm or `pilot wave' model of quantum mechanics~\cite{bohm1952suggested,albert1994bohm} describes only part of the motion of sonons because it omits the carrier waves.

Suppose, conterfactually, that a sonon changes state and becomes de-localized in some unspecified way, and then suddenly rematerializes just before any measurement, without affecting the trajectory. It follows from the above that the probability of rematerialization at any given position must be $|\psi|^2$. This is analogous to the Copenhagen interpretation of quantum mechanics, where the wavefunction and its collapse obey identical equations.

The foregoing can be described in the terminology of telecommunications theory. At velocities significantly smaller than $c$, the wavelength of $\chi$ (the {\em carrier wave}) is significantly shorter than that of $\psi$ (the {\em modulation}). Note that $\psi$ must be complex-valued to describe both amplitude and phase.

A similar emergence of quantum phenomena from completely classical motion is exhibited by small droplets which are made to bounce on a bath of the same liquid by oscillating the container vertically~\cite{couder2005dynamical}. The droplet's horizontal motion is determined by nonlinear interactions with the surface waves, which obey the wave equation and are a two-dimensional analogue of the disturbance $\xi = \chi \psi$. The principal differences are that the effective speed $c$ might be modified near the bouncing droplet due to the mass of the droplet and amplitude-related effects, and the experiment is a driven dissipative system rather than a lossless one. The videos show the droplets maintaining phase coherence with the surrounding waves due to nonlinearities. Single-slit diffraction, double-slit diffraction, unpredictable tunnelling and quantized energy levels are observed~\cite{couder2006single,protiere2006particle,eddi2009unpredictable,fort2010path}.

A sonon may be modelled as a classical oscillator of high Q. In interactions between such oscillators, Mead shows that the sum of the frequencies is conserved, since otherwise the interaction would average out to zero~\cite{Mead}. It follows from the definition $E = \hbar \omega$ (note, $E$ is not the same as energy in the fluid motion $E_{fluid}$ referred to above) that $E$ is conserved. The conservation of momentum, defined by ${\bf p} = \hbar {\bf k}$, follows from the conservation of $E$ and Lorentz covariance.

It follows from the above that the equations of motion for a sonon are completely classical, but they reduce to the established equations for a relativistic quantum mechanical particle of mass $\frac{\hbar \omega}{c^2}$ if the carrier wave is neglected.

\section{Long range force}
\label{sec:em}

In the bouncing droplet experiments~\cite{protiere2006particle}, the edge of the container can be modelled as an image droplet bouncing antiphase, leaving the meniscus undisturbed. Stroboscopic photographs show a droplet moves in a straight line until it approaches the edge, when it is deflected away. We interpret this to indicate a repulsive force from the antiphase image. Conversely, in-phase droplets form crystalline lattices, indicating an attractive force.

A similar phenomenon in three dimensions is used for degassing oils by subjection to ultrasonic vibration. As the bubbles expand and contract in response to the pressure waves, they induce flows in the fluid. When two bubbles are in-phase, no fluid will cross the mirror line between them, from symmetry, but the flows parallel to it reinforce, resulting in a reduced Bernoulli pressure and a force of attraction. The bubbles consequently merge and rise to the surface. 

The calculation of the force, adapted from Faber's textbook~\cite[\S 4.4]{Faber}, is as follows. If a vacuum cleaner hose operates in the wind, it will ingest momentum along with the air particles, and experience a force $F = \rho_o U Q_1$ where $Q_1$ is the flow and $U$ the wind velocity. If the wind is due to a similar hose with flow $Q_2$, then $U = Q_2 /(4 \pi r^2)$, giving an inverse square force of attraction
\begin{equation}
F ~~=~~ -\rho_o ~\frac{Q_1 Q_2} {4 \pi r^2}
\label{eq:fluid-force}
\end{equation}

The direction of flow, and of the force on both hoses, will be reversed if one of them is set to blow (with a baffle for spherical symmetry). If both are set to blow they will attract again. More generally, oscillatory motion results in an attractive force if it is in-phase, and a repulsion if it is antiphase. The phase alignment is most easily assessed from the flow across the mirror line between the sources.  

Turning to $R_{11}$ sonons, their flows are shown in Figure \ref{fig:r11fluidflow-with-cylinder} and their preferred alignments in Figure \ref{fig:spin-align}. From the mirror symmetry in \ref{fig:spin-align}(a) there is no flow across the mirror line and consequently sonons of opposite chiralities will attract one another. In \ref{fig:spin-align}(b) the direction of flow is reversed and like chiralites repel. The magnitude of the force is calculated in the appendix by substituting the idealised fluid flows of Figure \ref{fig:r11fluidflow-with-cylinder}(b) into (\ref{eq:fluid-force}) and calculating the cylinder's acceleration (a calculation which is only approximate)
\begin{equation}
	F \approx \frac{a^2 \beta^2}{64 b } ~C ~\sin^2(C)  ~\frac{\hbar c}{r^2}
\label{eq:coulomb-force} 
\end{equation}
where $a$, $b$ and $\beta$ are defined in Figure \ref{fig:r11fluidflow-with-cylinder}, and $C = \sqrt{k_r^2 R_o^2+1}$. Substituting the estimates $a \approx b \approx \beta \approx 1$ and $k_r R_o \lesssim 0.92$ from (\ref{eq:R11-radius}) gives
\begin{equation}
F~~\lesssim~~\frac{1}{49}~ \frac{\hbar c}{r^2}
\label{eq:fine-structure-constant}
\end{equation}
This inverse square force among $R_{11}$ sonons and their chiral twins is analogous to the Coulomb force among electrons and positrons. The fine structure constant $\alpha \lesssim 1/49$ matches the electromagnetic value to the accuracy of calculation.

\subsection{Radiation}
\label{radiation}

The interaction in \eqref{eq:fine-structure-constant} can be generalized by applying a Lorentz boost. We can borrow from the usual extension of the Coulomb force by requiring it to be Lorentz covariant, which produces Maxwell's equations, as shown in the appendix.

Maxwell's equations have solutions corresponding to propagating waves. We now show these waves are emitted when a sonon is accelerated. Without loss of generality we will examine one component of $\Phi$ in \eqref{eq:phi-r}.

If a sonon is made to oscillate so its position is $x = x_o \sin(\Omega t)$, where $x_o$ is small, then substituting into \eqref{eq:factorise-xi} gives (see the appendix)
\[
\xi ~~=~~ \psi_w \chi_w
  \]
where
\[
\psi_w ~~ = ~~ \frac{A R_o}{r} ~~ k_r x_o ~~ \sin \left[ \Omega \left(t - \frac{r}{c} \right) \right]
  \]
\begin{equation}
	\chi_w ~~= ~~B_1 ~ ~e^{-i \omega_o t} ~\cos(k_r r) ~ \cos \Theta
\label{eq:chi_w}
\end{equation} 
and $\Theta$ is the angle from the direction of oscillation. Here $\psi_w$ describes amplitude modulation of a carrier wave $\chi_w$, but in general there will be phase modulation as well for larger amplitude motion.

The momentum of the wave can be calculated by writing \eqref{eq:chi_w} as a sum of propagating waves
\begin{equation}
\chi_w = \frac{i B_1}{2} ~ 
		   [ e^{-i \omega_o (t + r/c)} 
		~ +  ~e^{-i \omega_o (t -r/c)} 
		   ]
\label{eq:inoutwave}
\end{equation}

These propagating solutions can be extended all the way to the particle (more precisely using a sum of Hankel functions). Extended in this way, they comprise the entire solution. Their sum must have the same energy and momentum as the particle, at all velocities, which enables a relationship between the energy $E_w$ and momentum $p_w$ of the waves to be calculated
\[ 
	p_w ~ = ~ \frac{E_w}{c}
\]
See the appendix for the calculation. This also applies to the radiating waves described above, which are a superposition of such waves. Notice the analogue with light, which carries the same momentum.

\subsection{Constructive interference}
\label{ss:bosons}

In very shallow water on a sandy beach, the wave speed increases with depth, and a wave travelling close behind another will propagate in the deeper water of the wave in front and travel faster, catching it up~\cite{Faber}.

A similar effect will occur in the waves under consideration due to the speed $c = \sqrt{\partial P/\partial \rho}$ varying over the cycle. The trailing wave will be perturbed by the wave in front, which remains in phase over a long time, and therefore it does not cancel over a cycle. This effect is not an ordinary expectation value (as defined in \S \ref{sec:lorentz-covariance}) and therefore it is not precisely Lorentz covariant.

There is an analogy with light waves, which also tend to come into coherence, such as in lasers. The underlying components of the waves in (\ref{eq:inoutwave}) will interfere constructively if one overtakes the other by about half a wavelength of the carrier wave. For comparison, an oscillating electron would have about $10^{12}$ half-wavelengths per metre in its carrier wave. On a crude model, two waves of modulation will reach coherence after about $1m$ if the speed difference is $\Delta c \sim 10^{-12} c$, which corresponds to about $0.0003m s^{-1}$.

\section{Causality and localization}

Euler's equation is deterministic, and therefore chance and dice cannot be involved at a fundamental level in the behaviour of sonons in a closed system. The difficulty of predicting outcomes arises, among other causes, from errors or uncertainties in the initial measurements, which are magnified in processes such as diffraction and tunnelling.

In a closed deterministic system, its state at some time $t_o$ completely dictates its state at some other time $t$. This is true even when $t_o>t$, which could be interpreted non-causally. Such solutions are valid because Euler's equation is symmetric under time reversal.

In practice, these coherent solutions are easily disrupted by influences from outside the closed system. In particular, the coherence between a sonon and its surrounding waves is fragile. The process of decoherence is most easily pictured in the droplet experiments, where a disturbance causes a droplet to bounce out of its local wave trough, magnifying the effect of the perturbation.

Nevertheless, the time-reversed motions are valid solutions if decoherence can be controlled. See Mead's analysis~\cite{Mead} from first principles on the interactions between classical oscillators through forward and time-reversed waves which are solutions to the (time-symmetric) wave equation. Interacting sonons obey similar equations. Provided phase coherence can be maintained during the transition, Mead shows that the interaction produces effects which are closely analogous to the quantisation of the photon, among other observed phenomena.

Mead's model is related to Cramer's more general transactional interpretation of quantum mechanics~\cite{cramer1986transactional}, which was designed to be consistent with experiments on Bell's inequality~\cite{bell1964einstein,aspect1982experimental,bell2004speakable} by exploiting the property that the equations are symmetric under time reversal. Cramer's model is completely local and time reversal symmetric, symmetries which are shared by Euler's equation. It might be possible to interpret the motion of sonons (still consistent with experiments on Bell's inequality) in a different way, without needing to invoke the time-reversal symmetry, by exploiting the fact that spin-related information is carried by the carrier waves. This transmission of information is not usually considered in the interpretation of the experiments. See ~\cite{Anderson2013quantum} for further discussion.

\section{Conclusion}
\label{sec:conclusion}

The irrotational motion of a compressible inviscid fluid is studied in the field of analogue gravity, where there are phenomena related to general relativity and the standard model~\cite{barcelo2011analogue,unruh1981experimental,lahav2010realization,Volovik-2003}. The main contribution of this paper has been to extend the analogy to quantum and electromagnetic phenomena. We introduced the sonon, a quasiparticle with a twist, which behaves like a relativistic quantum mechanical particle with spin-$\frac{1}{2}$ symmetry, and which experiences an interaction, due ultimately to Bernoulli forces, that obeys Maxwell's equations.

The number and precision of these analogies suggests there might be a wider explanation. One candidate arises from an argument advanced by Einstein in 1920~\cite{einstein2007ether}. Unlike linear motion, acceleration and rotation are intrinsically discoverable. For example, a gyroscope's axis tends to remain aligned with the distant stars. Believing all interactions to be local, Einstein reasoned these correlations between distant objects must be mediated by a substance occupying the space between them, which he called a ``medium for the effects of inertia'' or ``ether of the general theory of relativity''. By requiring the medium to be consistent with special relativity he deduced that it cannot have any mechanical properties if it is a solid or a semi-solid.

A fluid medium was not considered at the time because it cannot support transverse waves, which were thought to be needed to model polarised light. However, this assumption cannot be justified because a compressible inviscid fluid supports, not transverse waves, but waves of modulation which obey Maxwell's equations.

These observations are consistent with the hypothesis, which we now advance, that Einstein's inertial medium behaves as a nonrelativistic barotropically compressible inviscid fluid. The hypothesis is testable because it predicts all observable quantities are completely determined by the solutions to Euler's equation for such a medium.

The observable -- and observed -- quantities described above in general relativity, particle symmetries, quantum mechanics and electromagnetism can be derived from this single hypothesis. In addition, it predicts that all ordinary expectation values (as defined in \S \ref{sec:lorentz-covariance}) must be Lorentz covariant to a very good approximation at all amplitudes. This is consistent with the original paper on special relativity~\cite{einstein1905electrodynamics}, which specifies that the postulates apply to the speed of light in empty space and the phenomena of electrodynamics and mechanics (we interpret this to refer to the corresponding ordinary expectation values). However, our hypothesis is not consistent with a possible extension to the original postulates, in which all motion is presumed to be Lorentz covariant. In particular, there are rare expectation values which do not average over a cycle but still have observable consequences. These are not precisely Lorentz covariant and, as described in \S \ref{ss:bosons}, such deviations contribute to, or may be responsible for, the coherence effects observed in lasers.

The calculation of the fine structure constant might be improved by computer simulation and compared to the observed electromagnetic value. The families of quasiparticles and the redshift near a massive body in \S \ref{ss:standing-waves} might be studied further and related to analogue gravity.

I warmly thank Andrew McLachlan, Keith Moffatt, Graziano Brady, Peter Landrock, and especially Robin Ball and Ross Anderson for immensely helpful comment and discussion.
%
%\bibliography{sonon}
%\bibliographystyle{unsrt}

\newpage
\appendix

\ifappendixtitle
\else
	\appendixpage
\fi

\section{Symbols}

\begin{tabbing}
{\bf Symbol} ~~~~~\= {\bf Meaning} \\
\\
$({\bf x}, t)$ \> Cartesian coordinates\\
$(r, \theta, s)$ \> Cylindrical coordinates\\
$\phi$ \> Angle around torus\\
\\
$P({\bf x}, t)$ \> Fluid pressure\\
${\bf u}({\bf x}, t)$ \> Fluid velocity\\
$\rho({\bf x}, t)$ \> Fluid density\\
$\rho_o$ \> Mean fluid density\\
$\xi({\bf x}, t)$ \> Excess density\\
 \>defined by $\rho/\rho_o = 1 + Re(\xi)$\\
$c$ \> Speed of propagation\\
 \>defined by $c^2 = \partial P/\partial \rho$\\
\\
$J_m(r)$ \> Cylindrical Bessel function \\
 \>of the first kind\\
$j_m(r)$ \> Spherical Bessel function \\
 \>of the first kind\\
$R_{mn}$ \> Spatial dependence of a sonon\\
\\
$\bf v$ \> Velocity of Lorentz boost\\
$\gamma$ \> Lorentz factor $1 / \sqrt{1 - v^2/c^2}$\\
\\
$\psi({\bf x}, t)$ \> Copenhagen wavefunction\\
$\chi({\bf x}, t)$ \> Carrier wave $\xi = \chi \psi$\\
$(\omega, {\bf k})$ \> Angular frequency, wavevector\\
\end{tabbing}

\section{Flow pattern of an eddy}
\label{a:eddy-flows}

The article discusses a solution to Euler's equation at low amplitude, given by 
\ifnature
	\begin{equation}
		\xi ~ = A ~ e^{-i(\omega_o t + m \theta - k_s s)} ~J_m (k_r r)
	\label{eq:eddy}
	\end{equation}
\else
	\[
		\xi ~ = A ~ e^{-i(\omega_o t + m \theta - k_s s)} ~J_m (k_r r)
	\]
\fi
where here, and throughout this appendix, the symbols have the meanings in the table above.

At low amplitude the quadratic term in Euler's equation can be neglected, giving $\rho_o \partial{\bf u}/\partial t = -\nabla P$. Substituting the definition $c^2 = \partial P/\partial \rho$ gives $\rho_o \partial {\bf u}/\partial t = c^2 \nabla \rho$ or 

\begin{equation}
{\bf u} ~~ = ~~ \frac{c^2}{\rho_o} \int \nabla \rho ~ dt
\label{eq:eddy-flow}
\end{equation}

This flow is irrotational~\cite{Faber}, as can be seen by rearranging \eqref{eq:eddy-flow} into the form ${\bf u} = \nabla \phi$ where $\phi= c^2 \int s~ dt$ where $s=(\rho - \rho_o)/\rho_o$. 

\begin{figure}[htb]
	\centering
	\includegraphics[width=0.8\columnwidth]{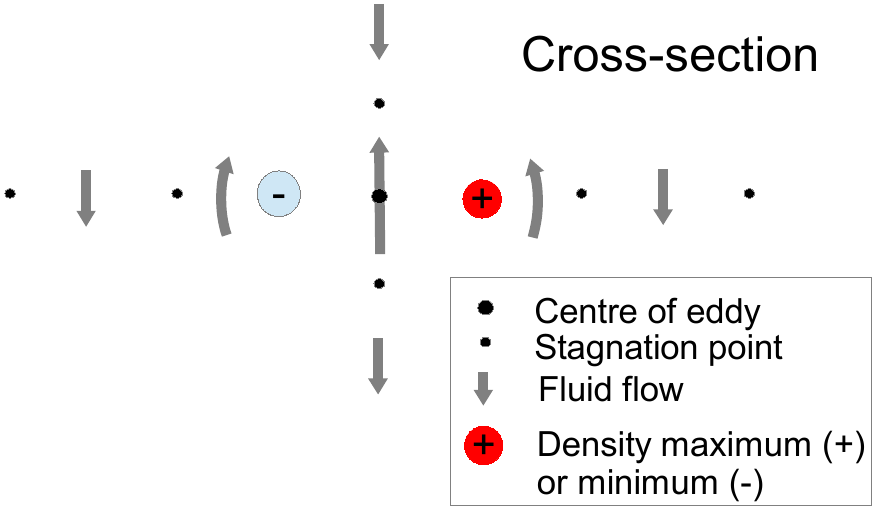}
	\caption{Flow pattern of an eddy (equation \ref{eq:eddy} where $m = 1$ and $k_s = 0$) at the instant $t = 0$}
	\label{fig:eddy-flow}
\end{figure}

Figure \ref{fig:eddy-flow} sketches the flow pattern near such an eddy with $m = 1$, which can be calculated by expressing \eqref{eq:eddy-flow} in cylindrical coordinates 

\[
u_r ~~ = ~~ \int ~ \frac{c^2}{\rho_o} ~ \frac{\partial \rho}{\partial r} ~ dt
  \]

\[
u_{\theta} ~~ = ~~ \int ~ \frac{c^2}{\rho_o r} ~ \frac{\partial \rho}{\partial \theta} ~ dt
  \]
  
Substituting into (\ref{eq:eddy}) and integrating gives

\[
	u_{r} ~~ = ~~ A ~ \frac{c^2}{\omega_o}  ~~ \frac{\partial}{\partial r} J_m ( k_r r) ~ \sin (\omega_o t + m \theta - k_s s) 
\]

\[
u_{\theta} ~~ = ~~ A ~ \frac{c^2}{\omega_o}  ~~ \frac{m}{r} ~ J_m ( k_r r) ~ \cos (\omega_o t + m \theta - k_s s)
\]

In Figure \ref{fig:eddy-flow} the $x$ axis is aligned with $\theta = 0$ and the flow pattern is sketched at the instant $t = s = 0$. The flow is stationary at the points marked with a small dot on the diagram. On the $x$-axis this stationary flow occurs when $J_1(k_r r) = 0$, or $k_r x  = \{3.83, 7.01, 10.71 ..\}$. On the y-axis it occurs when $(\partial/\partial r) J_1(k_r r) = 0$, or $k_r y = \{1.84, 5.33, 8.53 ..\}$.

\section{A sonon as a sum of spherical harmonics}
\label{a:sonon-spherical-harmonics}

A sonon is obtained by curving an eddy \eqref{eq:eddy} into a ring. This is a good description at small $r$, but the curvature of the $s$ axis makes the analysis more difficult at larger $r$. We now describe a solution to the wave equation which behaves in the same way at large distance, and uses more tractable coordinates.

\begin{figure}[htbp]
  \centering
	\includegraphics[width=0.2\textwidth]{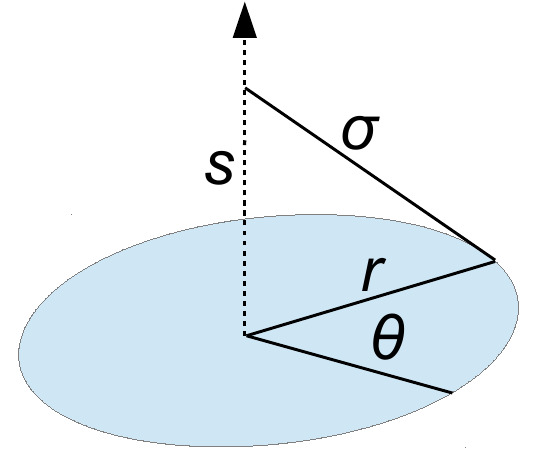}
	\caption{The coordinates used in \eqref{eq:cylindrical-spherical-bessel}}
	\label{fig:eddy-integration-path}
\end{figure}

In the coordinates of Figure \ref{fig:eddy-integration-path}, we begin by showing that

\begin{equation}
J_m(r) ~~ = ~~ \int_{-\infty}^\infty j_m(\sigma)~ ds ~~~~~~  (r \rightarrow \infty)
\label{eq:cylindrical-spherical-bessel}
\end{equation}

Expanding the spherical Bessel function at large distance, the dominant term is

\[
	j_m(\sigma)~~ = ~~ \frac{1}{\sigma} \sin \left(\sigma - m \frac{\pi}{2} \right) ~~~~~(r \rightarrow \infty)
\]

The integral in (\ref{eq:cylindrical-spherical-bessel}) can be evaluated by Fresnel's method, that is, using

\[
	\sigma ~~= ~~r\left( 1 + \frac{s^2}{2 r^2} \right) ~~~~(r \rightarrow \infty)
\]
Choosing $m = 0$, in the first instance, gives the Fresnel integral

\[
	\int j_m(\sigma)~ ds \approx \frac{1}{r}~ \int_{-\infty}^{\infty} \sin \left(r ~ + ~ \frac{s^2}{2 r}  \right) ds ~ (r \rightarrow \infty)
\]
which evaluates to 

\[
	\int j_m(\sigma)~ ds \approx  -\sqrt{\frac{2}{\pi r}} ~~ \cos \left(r + \frac{\pi}{4}    \right) ~~~~ (r \rightarrow \infty)
\]
which is the same as the largest term in the expansion of the cylindrical Bessel function. It may be verified that this is true for all values of $m$.

This demonstrates \eqref{eq:cylindrical-spherical-bessel}. Substituting into the equation (\ref{eq:eddy}) for an eddy gives
\begin{equation}
	\xi ~~=~~ \int_{-\infty}^{\infty} A e^{-i(\omega_o t + m \theta - k_s s)} j_m (k_r \sigma) ~ k_r ds
	\label{eq:eddy-integral}
\end{equation}
The substitution is valid in the case $k_s=0$, and the equivalence of the two expressions for $k_s \neq 0$ follows by applying a Lorentz boost. 

When $m=0$ or $1$, which are the cases of interest in the article, the Legendre polynomial $P_m$ reduces to $P_m(cos \theta) = cos(m \theta)$ and the integrand is a solution to the wave equation. For larger values of $m$ an expansion in terms of Legendre polynomials is required.

Given that the integrand is a solution to the wave equation, the path of integration need not be restricted to a straight line. Any path of integration, corresponding to any curving of the eddy, can be taken, and the result will still be a solution to the wave equation (strictly speaking, an approximation is involved if $k_s \neq 0$ because we have used a Lorentz boost in the above derivation, which is restricted to a straight line, but we will neglect the perturbation). 

\ifnature
	\begin{figure}[htbp]
	  \centering
		\includegraphics[width=0.5\columnwidth]{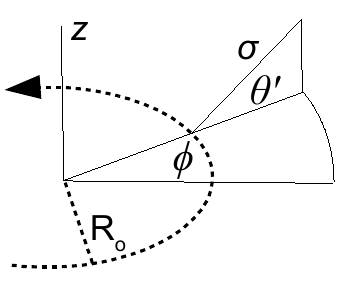}
		\caption{The coordinates used \eqref{eq:rmn}}
		\label{fig:sonon-integration-path}
	\end{figure}
\fi

In the coordinates of Figure \ref{fig:sonon-integration-path}, replacing $k_s s$ in (\ref{eq:eddy-integral}) by $n \phi$ because the motion must be single-valued gives
\[
%\begin{equation}
	\xi ~~=~~ A ~e^{-i \omega_o t} ~R_{mn}
	%\label{eq:sonon}
%\end{equation}
\]

\ifnature
	\begin{equation}
		R_{mn} ~~\approx ~~ \int e^{-i (m \theta' - n \phi)} j_m (k_r \sigma) ~R_o d\phi
		\label{eq:rmn}
	\end{equation}
\else
	\[
		R_{mn} ~~\approx ~~ \int e^{-i (m \theta' - n \phi)} j_m (k_r \sigma) ~R_o d\phi
	\]
\fi
which are the equations used in the article.

\section{Spin symmetry}

The main text shows that the density pattern associated with a R$_{11}$ sonon has the symmetry of the Pauli matrices. A possible physical interpretation of the associated spin-half symmetry is as follows. 

\begin{figure}[htb]
	\centering
		\includegraphics[width=0.50\columnwidth]{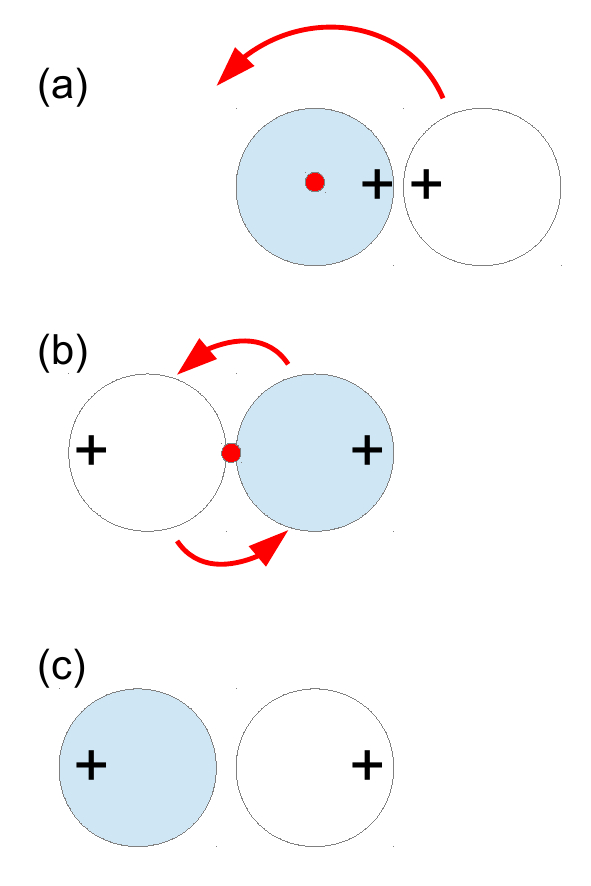}
	\caption{Two $R_{11}$ sonons maintaining their preferred alignment (with an antinode between them), drawn as if they were close to one another for clarity. A complete rotation is produced by the following idealised steps: (a) the blue sonon remains stationary whilst the white one rolls around it by half a turn, as if locked by gears, producing (b). The entire structure then rotates rigidly through a further half turn, producing (c). Observe the sign is reversed after what is effectively a complete rotation.}
	\label{fig:spin-half}
\end{figure}

It is shown in the main text that two identical sonons have a preferred alignment where there is an antinode of density between them and their spins are opposed. Refer to Figure \ref{fig:spin-half}, which considers the effect of rotating such a pair in space whilst maintaining this alignment. It may be noted that one complete rotation as drawn reverses the sign of the density pattern. This symmetry may be relatred to the spin-half symmetry associated with the Pauli matrices.

\section{Dirac equation}
 
The expressions related to the Pauli matrices in the main text can be written in Cartesian coordinates at large $r$
\begin{equation}
	i~ \sigma_j~.~\frac{\partial \Phi}{\partial x_j} ~~ =~~ \pm k_r \Phi 
\label{eq:dirac-o}
\end{equation}
where $\sigma_j$ are the Pauli matrices and summation over $j$ is implied. The contributions from the angular components have been neglected because of the term $\frac{1}{r}$ which appears when transforming the angular components from spherical to Cartesian coordinates. 

Notice that \eqref{eq:dirac-o} is analogous to the Dirac equation in the case of a stationary particle, where the time derivative is assumed to vanish. 

The main text shows that the irrotational solutions to Euler's equation are Lorentz covariant to a very good approximation at all amplitudes. Consequently \eqref{eq:dirac-o} can be generalised to an equation which applies at all velocities, by writing it in a Lorentz covariant form (using the procedure which is shown in detail in the main text for a scalar field). It may be noted that there is an analogue with the Dirac equation, which is also Lorentz covariant and reduces to \eqref{eq:dirac-o} when stationary.

\section{Fluid energy near a sonon}
\label{a:fluid-energy}

Here we obtain an approximate expression for the kinetic and potential energy in the fluid motion within a sphere of radius $R$ of a sonon. 

In the main text it is shown that the density pattern at large distance from a sonon is given approximately by
\[
	R_{mn} ~~\approx~~ R_o B.\Phi
\]
where $B$ is of order 1. Here we will calculate the energy for the term $\Phi = \sin(k_r r)/r$ (the term in $\cos(k_r r)/r$ behaves the same), replacing the constants which are of order 1, $A$ and $B$, simply by 1.

The amplitude of these standing waves is $\Delta \rho = \rho_o R_o/r$ at their peak. Substituting into the fluid speed~\cite{Faber}, which is $c \Delta \rho/\rho_o$, gives
\[
	u_{peak} ~~\approx~~ \frac{R_o}{r}~c
\]

The kinetic energy density $\rho_o u^2/2$ is equalled by the potential energy density when averaged over a cycle, and so the total energy inside a sphere of radius $R$ is

\[ 
	E_{fluid} ~~=~~ \int_0^R 4 \pi r^2 \rho_o u^2 dr
\]
The spatial integral gives a factor of 1/2, and averaging over time gives another factor of 1/2 giving
\ifnature
	\begin{equation}
		E_{fluid} ~~ \approx ~~ \pi R_o^2 ~ R~ \rho_o c^2
	\label{eq:sonon-energy-r}
	\end{equation}
\else
	\[
		E_{fluid} ~~ \approx ~~ \pi R_o^2 ~ R~ \rho_o c^2
	\]
\fi
which is the equation used in the main text.

\section{Boundary condition: coupling mechanism}
\label{sec:boundary-condition-coupling-mechanism}

In the main text, it is shown that the energy in the waves surrounding a sonon is given by \eqref{eq:sonon-energy-r}, which will be minimised if sonons align themselves so their waves interfere destructively at large distance. Here we describe a coupling mechanism which allows this lowest energy state to be reached. Huygens discovered a similar mechanism in pendulum clocks in 1665~\cite{bennett2002huygens}.

Two similar sonons will behave as oscillators which are  weakly coupled through he nonlinear term in Euler's equation. They will consequently have two modes of coupled oscillation, in-phase and antiphase, of frequencies 

\begin{equation}
f~~=~~f_o \pm \nu
\label{eq:frequency-perturbation}
\end{equation}
where $\nu$ depends on the strength of the coupling. 

If both modes in \eqref{eq:frequency-perturbation} are present, there will be a component of motion at the difference frequency $2 \nu$, which will be coupled to longitudinal waves by the nonlinear term in Euler's equation. The lower energy state can be reached by these waves carrying energy away, which will be favoured because of entropy considerations (assuming a low temperature at large distance). The mechanism will be efficient because the waves have a frequency which is significantly different from the natural frequency $f_o$ of the nearby sonons of a given type, and so they will have a long mean free path.

Notice that this mechanism will also have the effect of locking together the frequencies of the sonons of similar types in a local region, even if they initially have slightly different sizes and frequencies.

\section{Redshift near a massive body}
\label{a:redshift}

In the main text it is shown that the sonons near a massive body become enlarged and their frequencies redshifted. These are still solutions to Euler's equation, since if $\xi({\bf x}, t)$ is a solution then so is  $\xi(a {\bf x}, a t)$ where $a$ is a scale factor. In units where $a = 1$ at large distance, then $a < 1$ near the body. 

Here we outline a possible relationship between this redshift and the gravitational forces which are studied in analogue gravity. The treatment here uses the analogue with quantum mechanical and classical motion which is set out in the main text, and it is only intended to apply at large distance.

In a uniform situation, the frequency $f$ of the sonons of a particular type at any given position will be the average of the frequencies in the vicinity, due to transport processes, giving $\nabla^2 f = 0$. This applies to all types of sonons and therefore

\[
\nabla^2 a ~~=~~0
\]

The spherically symmetric solution to this equation can be written in the form

\[
	a~~=~~1 - \frac{2 r_o}{r}
\]
where $r_o$ is to be found. 

A test particle will be perturbed by this field, since its frequency will be redshifted by $a$ due to the Huygens correlation discussed in \S \ref{sec:boundary-condition-coupling-mechanism}. This corresponds to the relationship

\[
	\omega ~~=~~ \omega_o \left( 1 - \frac{2 r_o}{r} \right)
\] 

and multiplying by $\hbar$ gives an equivalent expression for the energy perturbation, which is propotional to the natural frequency or inertial mass of the sonon and inversely proportional to distance.

At large distance, this perturbation corresponds to an inverse square force proportional to the inertial mass of the test particle. From the conservation of momentum there must be an equal force on the attracting object, and, from symmetry, it must be proportional to the product of the inertial masses. This is analogous to Newton's law of gravitation. The effects of the redshift have been neglected and consequently this is only true at large distance.

%                                               ----------- Coulomb force ---------------
\section{Analogue of the Coulomb force}
\label{a:fsc}

Here we calculate the fluid dynamical force between two stationary $R_{11}$ sonons when they are in their preferred alignment shown in Figure \ref{fig:spin-align}.
\ifnature
	\begin{figure}[htb]
		\centering
			\includegraphics[width=.85\columnwidth]{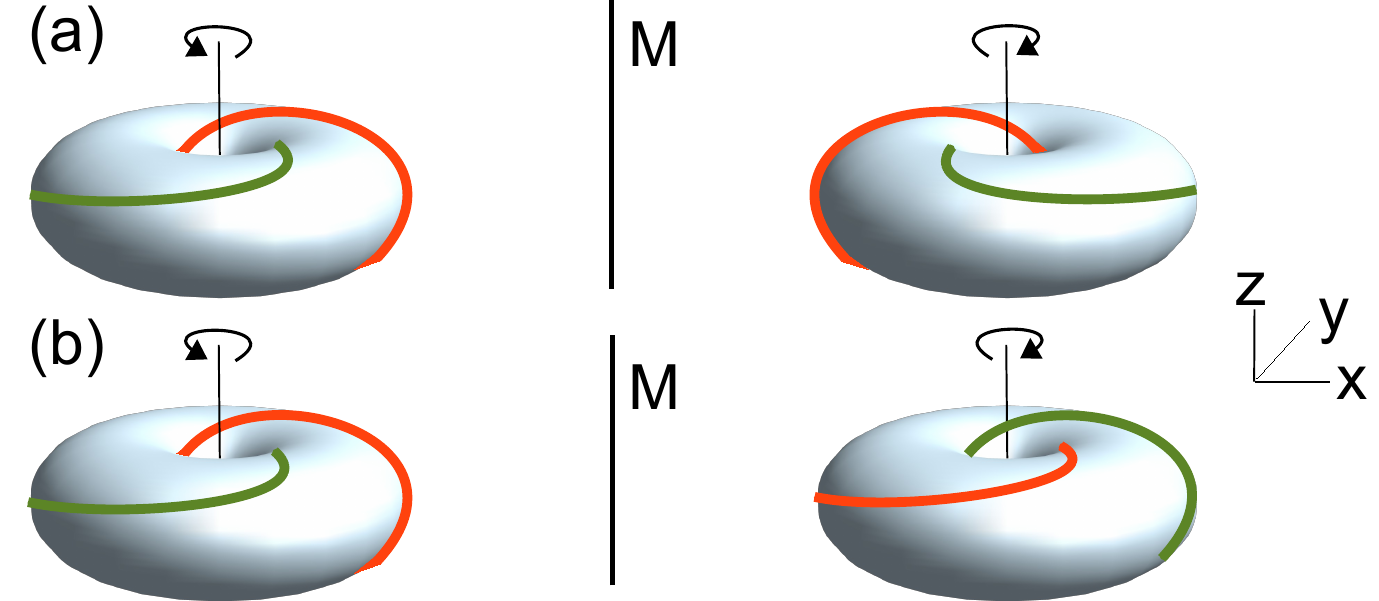}	\caption{The lowest energy alignment of $R_{11}$ sonons (a) opposite chirality (b) same chirality. The waves interfere destructively at large distance and constructively between them. There is an antinode of density on the mirror line $M$ and the spin vectors are opposed.}
		\label{fig:spin-align}
	\end{figure}
\fi

We will also use the calculation of the fluid dynamical force between two identical sources or sinks of magnitude $Q$ which are distance $r$ apart, which is found in Faber's textbook~\cite[\S4.5]{Faber}:-

\ifnature
	\begin{equation}
		F ~~=~~ \rho_o ~\frac{Q^2} {4 \pi r^2}
	\label{eq:fluid-force}
	\end{equation}  
\else
	\[
		F ~~=~~ \rho_o ~\frac{Q^2} {4 \pi r^2}
	\]
\fi
The flow pattern near a $R_{11}$ sonon is sketched in Figure \ref{fig:r11fluidflow-with-cylinder}(a). There is a region where the flows reinforce and the nonlinearity is significant. This will be modelled as a cylinder as shown in Figure \ref{fig:r11fluidflow-with-cylinder}(b). Density variations are rolled up into the effective mean flow speed.
\ifnature
	\begin{figure}[htb]
		\centering
			\includegraphics[width=\columnwidth]{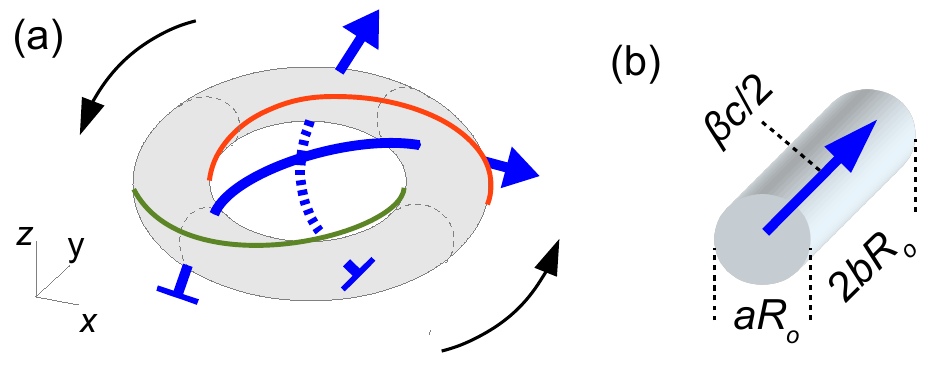}
		\caption{(a) Sketch of the flow near a $R_{11}$ sonon (b) The flows reinforce in a small region, which we idealise as a cylinder of diameter $a R_o$, length $2 b R_o$, and mean flow speed $\beta c/2$. A reasonable estimate is $a \approx b \approx \beta \approx 1$}
		\label{fig:r11fluidflow-with-cylinder}
	\end{figure}
\fi

In this idealisation, the external flow derives from a source and a sink at the ends of the cylinder in Figure \ref{fig:r11fluidflow-with-cylinder}(b), each of magnitude
\begin{equation}
	Q_i ~~ = ~~ \pm \pi \left( \frac{a R_o}{2} \right)^2 \frac{\beta c}{2}
\label{eq:cylinder-flow}
\end{equation}

These sources and sinks rotate with the sonon at angular speed $\omega_o$.  When viewed from large distance in the $x$ direction the flows cancel one another at the instant shown, as may be seen from the symmetry in Figure \ref{fig:r11fluidflow-with-cylinder}. However, when viewed from large positive $y$ there will be incomplete cancellation at the instant shown because of the time delays associated with information crossing the structure as it rotates. The incomplete cancellation is quantified by the approximate factor
\[
	  \cos \left(\omega_o \left[ t - \frac{b R_o}{c} \right] \right) 
	- \cos \left(\omega_o \left[ t + \frac{b R_o}{c} \right] \right) 
\]
and the resulting effective source becomes
\begin{equation}
	Q =~\frac{\pi a^2 R_o^2}{8} ~ \beta c ~~2 \sin(\omega_o t) ~\sin(\omega_o b R_o/c) 
\label{eq:R11-source-equation}
\end{equation}
Substituting into \eqref{eq:fluid-force} gives the force between these sonons
\[
	F = \frac{\rho_o}{64 r^2} \pi a^4 R_o^4 \beta^2 c^2 ~\sin^2(\omega_o t) ~\sin^2 \left(\frac{b ~\omega_o R_o}{c} \right) 
\]
Now, $\sin^2(\omega_o t)$ averages to 1/2 over a cycle, and the volume of the cylinder is $V = \pi a^2 b R_o^3/2$, giving the mean force
\begin{equation}
	F = \frac{1}{64 r^2} ~\frac{a^2}{b} ~\rho_o V \beta^2 c^2 ~ R_o ~\sin^2 \left(\frac{b ~\omega_o R_o}{c} \right) 
\label{eq:Coulomb-F-mechanical-units}
\end{equation}

The force produces an acceleration $a = F/m$ where $m$ is the mass of the fluid in the cylinder (where the nonlinearities are concentrated), which can be estimated as $\rho_o V$ neglecting density fluctuations. In the corresponding calculation of the attraction between bubbles which are subjected to ultrasonic vibration, the motion of the displaced fluid and the associated dressed mass is also taken into account~\cite{Faber}. This would give an additional factor of order 1/2, but we neglect it for simplicity, and it is likely to compensate for our neglecting density fluctuations. Thus we will use the approximate value
\[
	a = \frac{F}{\rho_o V}
\]

Expressed in conventional units, $F_c = (\hbar \omega/c^2) a$. Substituting into \eqref{eq:Coulomb-F-mechanical-units} gives
\begin{equation}
	F_c = \frac{a^2 \beta^2}{64 b} ~\frac{\hbar \omega_o R_o}{r^2}  \sin^2 \left(\frac{b ~\omega_o R_o}{c} \right)
	\label{eq:Coulomb-force1}
\end{equation}

For a linear eddy, $\omega_o^2 = c^2(k_r^2 + k_s^2)$, as discussed in the main text. For a $R_{11}$ sonon, which has one twist around the torus, $k_s = 1/R_o$ approximately, and so
\[
	\omega_o R_o ~~ = ~~ c~\sqrt{k_r^2 R_o^2 + 1}
\]
Substituting into \eqref{eq:Coulomb-force1} gives
\ifnature
	\begin{equation}
		F_c = -\frac{a^2 \beta^2}{64 b } ~C ~\sin^2(C)  ~\frac{\hbar c}{r^2}
	\label{eq:coulomb-force} 
	\end{equation}
\else
	\[
		F_c = -\frac{a^2 \beta^2}{64 b } ~C ~\sin^2(C)  ~\frac{\hbar c}{r^2}
	\]
\fi
where $C = \sqrt{k_r^2 R_o^2 + 1}$. This is used in the main text, dropping the unnecessary suffix $c$.

\section{Analogue of Maxwell's equations}
\label{a:maxwell}

The inverse square force in (\ref{eq:coulomb-force}) has been calculated for  stationary quasiparticles. Here we extend it to moving quasiparticles by using the symmetry that the solutions are Lorentz covariant. Our calculation is analogous to the usual extension of the Coulomb force to all velocities, and it will be no surprise that the resulting equations are analogous to Maxwell's equations. Whilst this function is formal, this section may nevertheless be of interest because, along the way, it also obtains the Lagrangian from the properties of a sonon in a transparent way, and it makes explicit the analogue between the phase of a sonon and the quantum mechanical phase which is measured in a superconductor using the Josephson effect.

The force in (\ref{eq:coulomb-force}) is curl-free, so it can be written as a potential gradient, $F = q \nabla \Phi$. The quantity $q$ is analogous to  the charge on a test particle and $\Phi$ to the sum of the potentials due to all other charges. 

Using Newton's second law in the form 
\[
	-\nabla \omega ~~=~~ \frac{\partial \textbf{k}}{\partial t}
\]
gives $q \nabla \Phi = \hbar \partial k/\partial t = \hbar \nabla (\partial S/\partial t)$ where $S$ is the phase of the sonon, which integrates to
\begin{equation}
	\frac{\partial S}{\partial t} ~~ = ~~ \frac{q \Phi}{\hbar} ~~ + ~~ constant
\label{eq:josephson}
\end{equation}
This key equation is analogous to the phase evolution in a charged superconductor, which obeys the same equation with $q=-2e$ and which is measured experimentally using the Josephson effect

A sonon moving between two fixed events suffers a phase change which must be stationary with respect to path variations. There would be destructive interference if this were not so. In free space the phase change is $\Delta S = E_o \tau / \hbar$ where $\tau$ is the proper time measured in a co-moving frame. In a stationary frame, $\hbar \Delta S = \int (E_o + T) dt$, where $T = (\gamma - 1) E_o$ is the kinetic energy.

This motion will be perturbed by the interaction in (\ref{eq:josephson}). Defining the Lagrangian

\[L = T - q \Phi
  \]
then $\hbar \Delta S = \int (E_o + L) dt$. If this phase difference is stationary with respect to path variations, $\delta (\hbar \Delta S) = 0$, then the Euler-Lagrange relationship $d/dt (\partial L/\partial v_i) = \partial L/\partial x_i$ reduces to

\[
\frac{d \bf p}{dt} ~~ = ~~ - q \nabla \Phi
  \]
which recovers Newton's second law for a particle of charge $q$ near stationary electrostatic charges.

Equation (\ref{eq:josephson}) can be extended to a moving system by exploiting the symmetry that the solutions are Lorentz covariant. The Lorentz covariant 4-vector is
\[
\left(\frac{1}{c} \frac{\partial S}{\partial t}, \nabla S \right) ~~ = ~~ \frac{Q}{\hbar} \left( \frac{\Phi}{c}, {\bf A}  \right) ~ + ~ \Gamma
  \]

Notice that $\bf A$ is analogous to the magnetic vector potential and $\Gamma$ the gauge. In a coherent system, the phase change around a loop must be quantised, so that
\[
\oint \nabla S {\bf dl} ~~ = ~~ \frac{Q}{\hbar} \oint {\bf A dl} ~~ = ~~ 2 n \pi
  \]
where $n$ is an integer. This is analogous to the flux quantisation which is observed in a superconductor.

The force in (\ref{eq:coulomb-force}) is inverse square, and in general such forces obey, in suitable units,
\[
- \epsilon_o ~ \nabla^2 \Phi ~~ = ~~ \rho_c
  \]
where the charge density $\rho_c$ represents the sources. This equation can also be extended into a Lorentz covariant form in the usual way, giving
\[
\left( \nabla^2 - \frac{1}{c^2} \frac{\partial^2}{\partial t^2} \right) \Phi ~~ = ~~ - \frac{\rho_c}{\epsilon_o}
  \]
  
Likewise, extending the density to a 4-current gives, more generally,
\[
\left( \nabla^2 - \frac{1}{c^2} \frac{\partial^2}{\partial t^2} \right) {\bf A} ~~ = ~~ - \mu_o {\bf J}
  \]
where $\bf J$ is the current. These are analogous to Maxwell's equations for the electromagnetic field.

%                                                                  ------ emission of radiation --------
\section{Radiation from an oscillating sonon}
\label{a:radiation}

As discussed in the main text, the density pattern of a sonon at large distance is given, without loss of generality, by

\begin{equation}
	\xi ~~= ~~A B_1 ~e^{-i \omega_o t}~ \frac{R_o}{r} \sin(k_r r)
	\label{eq:sonon-infinity}
\end{equation}

If a sonon oscillates so its position is $x = x_o \sin(\Omega  t)$ where $x_o$ is small and its velocity of motion is much less than $c$, then this becomes

\[
\xi ~~= ~~A B_1 ~ e^{-i \omega_o t} ~\frac{ R_o}{r} ~ \sin[k_r (r + r') ]
	\]
where
\[
r' ~~ = ~~ x_o \cos(\Theta) ~ \sin \left[ \Omega \left(t - \frac{r}{c} \right) \right]
  \]
and $\Theta$ is the angle from the direction of oscillation. This may be written in the form $\xi = \chi_w \psi_w$ where for low amplitude motion, using the first term in the Taylor series expansion in $k_r x_o$
\[
\psi_w ~~ = ~~ \frac{A R_o}{r} ~~ k_r x_o ~~ \sin \left[ \Omega \left(t - \frac{r}{c} \right) \right]
  \]
\[
	\chi_w ~~= ~~B_1 ~ ~e^{-i \omega_o t} ~\cos(k_r r) ~ \cos \Theta
\]
Notice that we have chosen to distribute the factors so that $\psi_w$ is a solution to the wave equation in its own right. This distribution differs (on a technicality) from that used for a sonon. These are the equations used in the main text.

\section{Energy and momentum of a wave}
\label{a:wave-momentum}

This appendix calculates the ratio of energy to momentum for a wave associated with a particle.
 
When at rest, the energy of a particle is $E = \hbar \omega_o$ so each of the outgoing and incoming waves  must have energy $E_w = \hbar \omega_o / 2$. This equivalence also applies after a Lorentz boost. In the case of parallel motion, the frequencies of the outgoing and incoming waves are Doppler shifted to $\omega_o D$ and $\omega_o/D$ respectively, where the Doppler shift is

\[
D ~~ = ~~ \sqrt{\frac{c + v}{c - v}}
\]

The total energy in this case is $\hbar \omega_o (D + 1/D) / 2 = \gamma \hbar \omega_o$, which matches the energy of the moving particle. A similar calculation for perpendicular motion shows this equivalence to be independent of direction.

A similar consideration applies to momentum. Suppose the momentum of each wave is $p_w = \beta E_w = \beta \hbar \omega_o$ where $\beta$ is to be determined. When at rest, the waves have opposite momenta so the total momentum vanishes. A Lorentz boost changes this to $p = \beta \hbar \omega_o(D - 1/D) = 2 \beta E v$.  This must be the momentum of the particle, $p = E v/c^2$, and therefore $2 \beta = 1/c^2$ and

\[ p_w ~ = ~ \frac{E_w}{c} \]

\end{document}